\documentclass[]{spie}  %>>> use for US letter paper
\usepackage{amsmath,amsfonts,amssymb}
\usepackage{xcolor}
\usepackage{graphicx}
\usepackage[colorlinks=true, allcolors=blue]{hyperref}
\usepackage{booktabs}
\usepackage[square,numbers]{natbib}
\title{Real-time guidewire tracking and segmentation in intraoperative X-ray}

\author[a,b]{\textcolor{black}{Baochang Zhang}}
\author[a,b]{\textcolor{black}{Mai Bui}}
\author[c]{\textcolor{black}{Cheng Wang}}
\author[a,b]{\textcolor{black}{Felix Bourier}}
\author[a,b,d]{\textcolor{black}{Heribert Schunkert}}
\author[a,e]{\textcolor{black}{Nassir Navab}}
\affil[a]{\textcolor{black}{Technical University of Munich, Germany}}
\affil[b]{\textcolor{black}{German Heart Center Munich, Germany}}
\affil[c]{\textcolor{black}{ZhongDa Hospital Southeast University, China}}
\affil[d]{\textcolor{black}{DZHK (German Centre for Cardiovascular Research), Munich Heart Alliance, Munich, Germany}}
\affil[e]{\textcolor{black}{Munich Institute of Robotics and Machine Intelligence (MIRMI), Munich, Germany}}

\authorinfo {Further author information: (Send correspondence to Baochang Zhang)\\ Baochang Zhang: E-mail: baochang.zhang@tum.de\\}

\begin{document} 
\maketitle

\begin{abstract}
During endovascular interventions, physicians have to perform accurate and immediate operations based on the available real-time information, such as the shape and position of guidewires observed on the fluoroscopic images, haptic information and the patients' physiological signals. For this purpose, real-time and accurate guidewire segmentation and tracking can enhance the visualization of guidewires and provide visual feedback for physicians during the intervention as well as for robot-assisted interventions. Nevertheless, this task often comes with the challenge of elongated deformable structures that present themselves with low contrast in the noisy fluoroscopic image sequences. To address these issues, a two-stage deep learning framework for real-time guidewire segmentation and tracking is proposed. In the first stage, a Yolov5s detector is trained, using the original X-ray images as well as synthetic ones, which is employed to output the bounding boxes of possible target guidewires. More importantly, a refinement module based on spatiotemporal constraints is incorporated to robustly localize the guidewire and remove false detections. In the second stage, a novel and efficient network is proposed to segment the guidewire in each detected bounding box. The network contains two major modules, namely a hessian-based enhancement embedding module and a dual self-attention module. Quantitative and qualitative evaluations on clinical intra-operative images demonstrate that the proposed approach significantly outperforms our baselines as well as the current state of the art and, in comparison, shows higher robustness to low quality images.
\end{abstract}
% \mai{You are still mentioning the synthetic images in the abstract but it is never explained later. Also now the data augmentation mentioned for the detector's performance Tab. 3 is never explained}
\section{INTRODUCTION}
\label{sec:intro}  % \label{} allows reference to this section
% During percutaneous coronary interventions, guidewires, as one of the most important surgical instruments, are inserted into the patient's vascular system, and advanced to reach the target area under the guidance of real-time X-ray fluoroscopy to provide treatment such as stenting, ablation or drug delivery~\cite{baert2003guide}. The guidewire must be precisely manipulated to avoid damaging the vessel wall, which in turn could lead to perforation and hemorrhage, both of which are fatal.
Guidewires are essential tools used in every interventional cardiology procedure. They are advanced into the cardiovascular system using percutaneous vascular access into the target area of the intervention under the guidance of real-time X-Ray fluoroscopy. Once the initial guidewire placement is successfully achieved, other catheters are advanced over the guidewires in order to provide treatments, such as stenting, ablation or drug delivery~\cite{baert2003guide}. Therefore a precise navigation of the guidewire is mandatory in order to avoid damaging vascular structures, which would result in procedural complications such as perforations, dissection, stroke and hemorrhage. For this, a robust guidewire tracking and segmentation system to analyze the consecutively captured X-ray images can be of significant help and, in addition, reduce the mental workload required by the interventionist as well as prevent visual fatigue~\cite{mazomenos2016survey}. Moreover, such a system can provide visual feedback of the guidewire to robotic systems, paving the road for semi- or fully automatic robot-assisted interventions.

However, due to the following reasons, guidewire segmentation and tracking still remains a challenging task: (a) guidewires present themselves as elongated deformable structures with low contrast in noisy X-ray fluoroscopy images; (b) only a small part of the guidewire is visible in the image, i.e. only a 3cm part of the guidewire contains radiopaque material and (c) their visual appearance easily resembles other anatomical structures (such as rib outlines or small vessels) in the fluoroscopic images. %Guidewire tracking and segmentation still remains a challenging task. 

Nowadays, most researchers define this task directly as a segmentation task to extract the guidewire from individual X-ray images. In general, resulting methods can be divided into traditional methods and deep learning-based methods. For traditional methods, the guidewire is extracted based on primitive features such as texture, a histogram or pixel intensity values. Meanwhile, these methods rely on manual annotation of the guidewire in the first frame of the fluoroscopic video sequence and the guidewire between two consecutive frames should not be significantly deformed~\cite{chen2016guidewire}. In ~\cite{vandini2017robust}, segment-like features (SEGlets) were introduced to overcome large deformations between successive frames. However, hand-crafted features tend to have poor generalization capabilities and robustness, especially in noisy environments. Recently, deep learning has achieved promising results in medical image analysis. In ~\cite{wu2018automatic,li2019two}, instead of segmenting the guidewire directly~\cite{ambrosini2017fully,zhou2020real}, a detector is deployed first to output the bounding box of the guidewire, which could reduce the class imbalance of the segmentation task. However, these methods currently ignore the importance of temporal information between continuously recorded frames, which can easily result in a high false positive rate in the segmentation results. 

%Considering the high requirement on real-time, we pay more attention on deep learning methods. 
In this paper, a two-stage system is proposed for real-time guidewire tracking and segmentation, making the following contributions: (1) A bounding box detector for guidewire detection is trained, using the original X-ray images as well as synthetically generated ones, followed by a refinement module based on spatiotemporal constraints to robustly localize the guidewire and suppress false positive detections. (2) A novel efficient network for guidewire segmentation is proposed, which achieves promising performance and strong robustness to low quality X-ray images by incorporating a hessian-based enhancement and attention modules. (3) The whole system is designed to perform in real-time and achieves an inference rate of approximately 35 FPS on a GPU. 

\section{METHOD}
\label{sec:method}  % \label{} allows reference to this section
\begin{figure}[t!]
   \begin{center}
   \setlength{\abovecaptionskip}{0.cm}
   \includegraphics[width=0.9\textwidth]{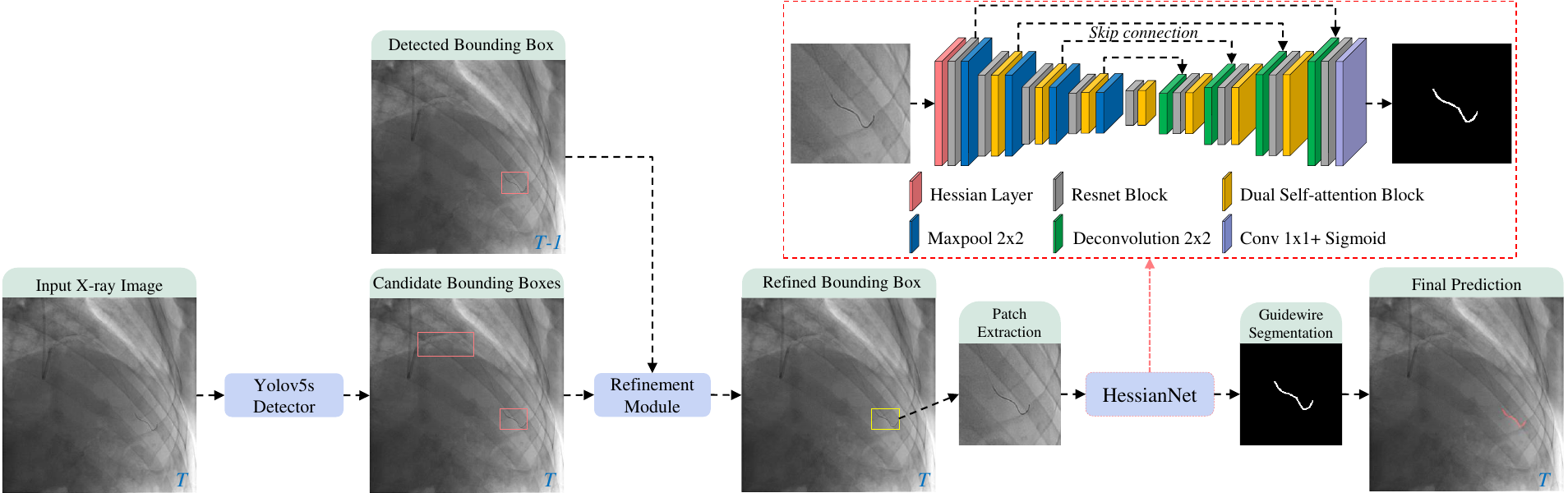}
   \end{center}
   \caption{ 
    Our framework consists of two stages. First, the guidewire is detected in the current image frame at time step $T$ by including temporal information and the detection of the previous frame at time step $T-1$. Second, the guidewire is segmented in the predicted bounding box by our proposed HessianNet, which, in addition to the image information, utilizes Hessian-based features for improved guidewire segmentation.}
   \label{fig:framework} 
   \vspace{1mm}
\end{figure}

In our work, we propose a two-stage framework for guidewire tracking (see Fig \ref{fig:framework}), that consists of 1) a temporal-aware detection stage and 2) a hessian-enhanced segmentation stage. 

% \noindent \textbf{Guidewire Localization.}
\subsection{Guidewire Localization.}
In the first stage, YOLOv5s is chosen as the detector to ensure real-time performance of the detector~\cite{jocher2020yolov5}, however, we make the following adjustments: 

(1) YOLOv5s only considers spatial but not temporal information. However, due to rather smooth motions we can assume the movement of the guidewire in the image space in consecutive frames too be limited. %; and existing guidewires do not suddenly disappear, and also suddenly appear where there was no guidewire before. 
Therefore, to leverage temporal information we use three consecutive frames as input and introduce a refinement module to further improve the detection precision.

(2) In YOLOv5s, each predicted bounding box is associated with a confidence score and a threshold manually set to obtain the final detection. However, an optimal threshold can be difficult to find and, in addition, easily result in false positives and false negatives in the detection results. To alleviate this issue, instead, the confidence threshold is determined automatically based on the maximum confidence value among all the detection boxes, e.g. $0.001 \cdot \text{\textit{maximum confidence value}}$. 

(3) In our method, the output bounding boxes of YOLOv5s are divided into two lists \textit{Pred}$_{H}$ and \textit{Pred}$_{L}$ depending on whether a box's confidence score is larger than 0.5, \textit{Pred}$_{H}$, or lower, \textit{Pred}$_{L}$. Since the position of the guidewire in two consecutive frames should be close, we calculate a matching relationship among the boxes in two consecutive frames by their Intersect over Union (IOU-score). Then, all the boxes, whose IOU-score with the box at timestep $t-1$ is larger than $\tau$=0.25, are merged as refined candidates of the bounding box at timestep $t$. Finally, the updated boxes are stored in \textit{$O^t$}; the unmatched boxes in \textit{$O^{t-1}$} and \textit{Pred}$^{t}_{H}$ are stored in \textit{$\hat{O}^t$}. Here, \textit{$O^t$} is used to store the bounding boxes already confirmed to contain a guidewire at timestep $t$; \textit{$\hat{O}^t$} is used to store the bounding boxes temporarily believed to contain a guidewire at timestep $t$ and need to be further confirmed at the next timestep. The complete process of refinement module is shown in 
Alg.1.

\begin{figure}[t!]
  \begin{center}
  \includegraphics[scale=0.85]{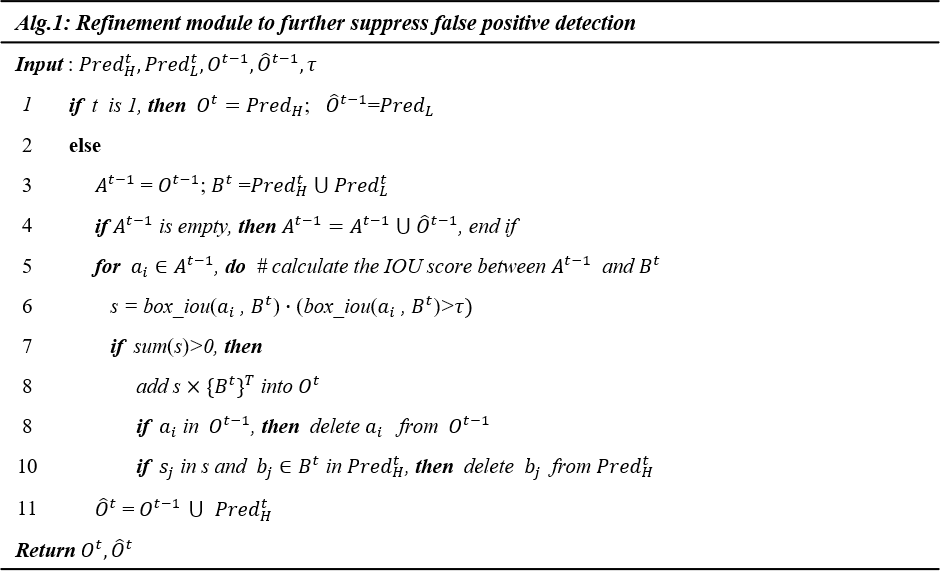}
  \end{center}
  \label{fig:Alg1} 
\end{figure} 

% \begin{algorithm}
% \caption{Refinement module to further suppress false positive detection}\label{alg:cap}
% \hspace*{\algorithmicindent} \textbf{Input: \textit{Pred}$^{t}_{H}$, \textit{Pred}$^{t}_{L}$, \textit{$O^{t-1}$}, \textit{$\hat{O}^{t-1}$}, \textit{$\tau$}}
% \begin{algorithmic}
% \If{ $t is 1$}
%     \State $O^{t} \gets Pred^{t}_{H}$
%     \State $\hat{O}^{t-1} \gets Pred^{t}_{L}$
% \Else
%     \State $A^{t-1} \gets O^{t-1}$ 
%     \State $B^{t} \gets Pred^{t}_{H} \cup Pred^{t}_{L}$
%     \If{$A^{t-1} is empty$}
%         \State $A^{t-1} \gets A^{t-1} \cup \hat{O}^{t-1}$
%     \EndIf
%     \For{$a_{i} \in A^{t-1}$} \Comment{calculate the IOU score between $A^{t-1}$ and $B^{t}$}
%         \State $s \gets box-iou(a_{i}, B^{t}) \cdot  (box-iou(a_{i}, B^{t}) > \tau)$
%         \If{$sum(s) > 0 $}
%             \State $add s \times {B^{t}}^{T} into O^{t}$
%             \If{$a_{i} \in O^{t-1}$}
%                 \State $delete a_{i} from  O^{t-1}$
%             \EndIf
%     \EndFor
% \end{algorithmic}
% \end{algorithm}

\subsection{Guidewire Segmentation.}
\begin{figure} [t!]
   \begin{center}
   \includegraphics[width=0.9\textwidth]{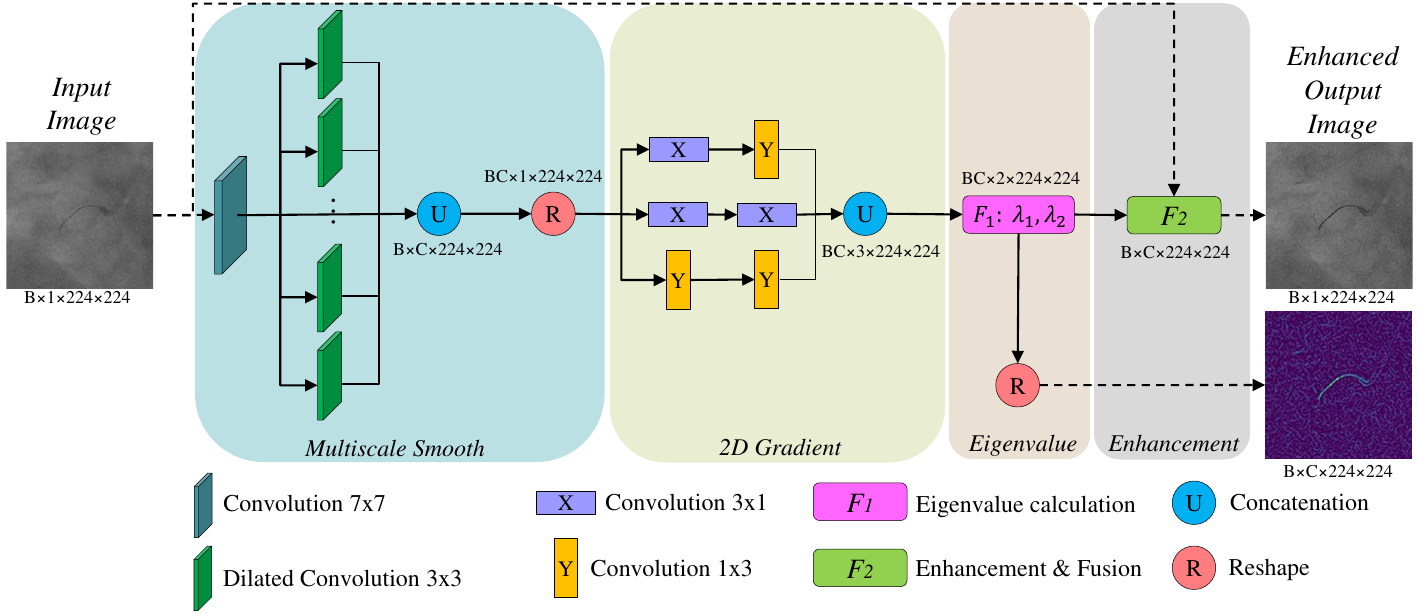}
   \end{center}
   \caption{ The architecture of the proposed Hessian Layer, which outputs multi-scale eigenvalue features $\lambda_{1}$ and enhances the guidewire in comparison to the original 2D images.}
   \label{fig:hessianlayer} 
   \vspace{2mm}
\end{figure}

After the detection stage, the patches that contain the guidewire are cropped based on the detected bounding boxes. Then, we propose HessianNet for guidewire segmentation, which consists of three modules: a hessian layer, an encoder module and decoder module, as shown in the red dotted box in Fig.~\ref{fig:framework}. Input grayscale patches of size $224 \times 224$ are first processed by the hessian layer, as shown in Fig.~\ref{fig:hessianlayer}, which is used to extract eigenvalue features and improve the quality of the input image. Next, the improved image is fed into the encoder module and the eigenvalue features are concatenated with the features extracted by the first Resnet-block. Maxpooling is then performed in the area of $2 \times 2$ with a stride of 2. For the following, each encoder block in the encoder consists of a Resnet-block, a dual self-attention block, which is used to capture the long-range dependency of the features, and a maxpooling layer. Then, the decoder module is employed to integrate the high-level and low level features and recover the resolution of the feature maps from $14 \times 14$ to $224 \times 224$. Each decoder block in the decoder consists of a deconvolution, batch normalization, a Resnet-block and a dual self-attention block. Finally, a convolution with a kernel size of $1 \times 1$ and a sigmoid function are used to output the segmentation result. 

\noindent \textbf{Hessian layer.} Within the hessian layer, we reformulate the hessian-based enhancement process as part of the neural network, such that it can be trained with our network architecture. First, a $7 \times 7$ convolution and several $3 \times 3$ dilated convolution operations are employed to extract multi-scale features, where the weights are initialized using the Gaussian function. Then, a $1 \times 3$ 1D-convolution and a $3 \times 1$ 1D-convolution are used to calculate the second-order gradients on each feature map, where the weight of the 1D-convolution is frozen as [0.5, 0, -0.5]. Later, the two eigenvalue maps $\lambda_{1}$ and $\lambda_{2}$ of each feature map can be calculated as:
\begin{equation}
    \lambda_{1}(D_{xx},D_{xy},D_{yy}) = \frac{(D_{xx}+D_{yy})-\sqrt{(D_{xx}-D_{yy})^{2}+4D_{xy}^{2}}}{2}
\end{equation}

\begin{equation}
    \lambda_{2}(D_{xx},D_{xy},D_{yy}) = \frac{(D_{xx}+D_{yy})+\sqrt{(D_{xx}-D_{yy})^{2}+4D_{xy}^{2}}}{2} 
\end{equation}

where $D_{xx},D_{xy},D_{yy}$ are the second-order gradients. In this work, the enhancement function proposed by \cite{jerman2016enhancement} is employed to produce the enhancement map. Finally, the input image and enhancement map are fused to improve the contrast of the guidewire. Meanwhile, the eigenvalue feature $\lambda_{2}$ is also stored and later reused to provide preliminary guidelines for guidewire identification.

\noindent \textbf{Dual self-attention block.} Note, instead of naively integrating the dual self-attention module at the end of the encoder \cite{mou2021cs2}, we employ it at each level of the encoder and decoder. However, the pair-wise attention computation among all pixels is highly inefficient and redundant. For improved efficiency, we first use a max-pooling operation to project the high-dimensional features to low-dimensional space and then calculate the low-rank spatial and channel affinity matrix. 

\section{Experiments}
\label{sec:results}  % \label{} allows reference to this section
In the following we evaluate our method with respect to both detection as well as segmentation performance and in addition evaluate the robustness of our and state-of-the-art methods with respect to the input image quality.
\subsection{Dataset}
102 sequences from 21 patients are collected from ZhongDa Hospital Southeast University, China, overall containing 5238 X-ray images with a resolution of $512 \times 512$. The images are divided into a training (4112 images from 80 sequences) and testing set (1126 images from 22 sequences). The ground truth of the guidewire in each frame has been annotated by an experienced intervention radiologist.
% In addition, we augment the training set by performing elastic deformations and synthetic image generation.

\subsection{Implementation Details.}
\begin{figure}[t!]
   \begin{center}
   \includegraphics[width=0.8\textwidth]{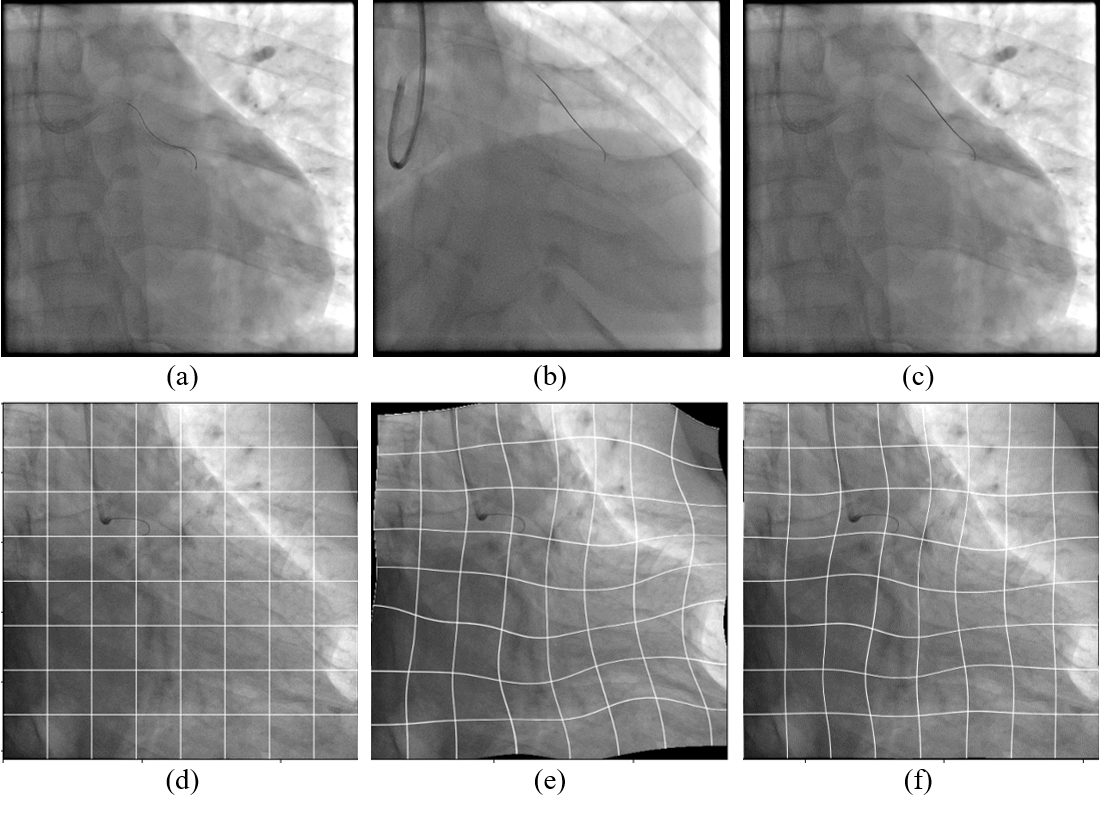}
   \end{center}
   \caption{ Exemplary synthetic images and elastic deformation. A synthetic image (c) is generated based on source images (a) and (b). A elastic deformed  image (e) and a boundary fixed elastic deformed  image (f) is generated based on source image (d).}
   \label{fig:aug} 
   \vspace{2mm}
\end{figure}

In this work, the Dice loss is adopted as the objective function for the training of the network. The proposed HessianNet was implemented in the PyTorch library with a  NVIDIA GPU (Quadro RTX 6000). Adam optimizer is employed as the overall optimizer where the learning rate is set as 0.0005 initially and decreases using exponential decay with the decay rate of 0.9. The batch size is set to 8, and the network is trained with a maximum number of 100 epochs. For the detection phase, the input image with its origin size of $512 \times 512$ is used and for data augmentation, boundary fixed elastic deformations and synthetic image generation is used. For that purpose, the synthetic images are generated by merging the guidewire from one sequence and the background from another X-ray sequence, as shown in Fig.~\ref{fig:aug}. For the segmentation phase, the size of the input image is cropped to $224 \times 224$, and for data augmentation, random horizontal flip, random vertical flip, random rotation [$-90,90$], random brightness ratio [0.8,1.2] and random contrast ratio [0.8, 1.2] are adopted.

\subsection{Guidewire Detection.}
Here, three simple evaluation metrics are used: to count the number of true positives (TP), false positives (FP) and false
negatives (FN) in all frames in the test sequences. TP indicates that the guidewire is completely within the predicted bounding box; FP indicates that there is no guidewire within the predicted bounding box; FN means that the guidewire is not detected or the predicted bounding box cannot cover the guidewire completely.

Table~\ref{table:detection} summarizes the performance of our detector where we compare between the original Yolov5s model, the inclusion of 1) data augmentation during training (in the form of elastic deformation and synthetic image generation, i.e. fusing the guidewire of one sequence with the background of another X-ray sequence), 2) three consecutive frames as input to the detector (C3) and 3) the proposed bounding box refinement step. From the Table~\ref{table:detection}, we can see that the original Yolov5s works well on this task, but there are still some targets missing and some false positive results. Our improved model (i.e. Yolov5s+C3+Aug.+Ref.), in comparison, can significantly reduce the number of missing targets and false positives in the outputs.

\begin{table}[t!]
	\begin{center}
		\caption{Inclusion of temporal information and the detector's performance. Our refinement step is abbreviated as Ref., Aug. and C3 stand for the inclusion of data augmentation and consecutive frames as input to the detector.}
		\vspace{2mm}
		\smallskip
		\footnotesize
		%\resizebox{\textwidth}{!}{	
		\setlength{\tabcolsep}{6pt}
			\begin{tabular}{lcccc}
			    \noalign{\smallskip}
			    \toprule
			      & Yolov5s & \vtop{\hbox{\strut Yolov5s }\hbox{\strut Aug.}} & \vtop{\hbox{\strut Yolov5s }\hbox{\strut C3+Aug.}} & \vtop{\hbox{\strut Yolov5s }\hbox{\strut C3+Aug.+Ref.}}\\
				\noalign{\smallskip}
				\midrule
				 TP & 1082 & 1109 & 1112 & 1125\\
				 FN & 44 & 17 & 14 & 1\\
				 FP & 7 & 2 & 1 &0 \\
				 \noalign{\smallskip}
				\bottomrule
			\end{tabular}
		%}
		\label{table:detection}
	\end{center}
	\vspace{1mm}
\end{table}

\subsection{Guidewire Segmentation.}
\begin{figure}[t!]
   \begin{center}
   \vspace{-2mm}
   \includegraphics[width=\textwidth]{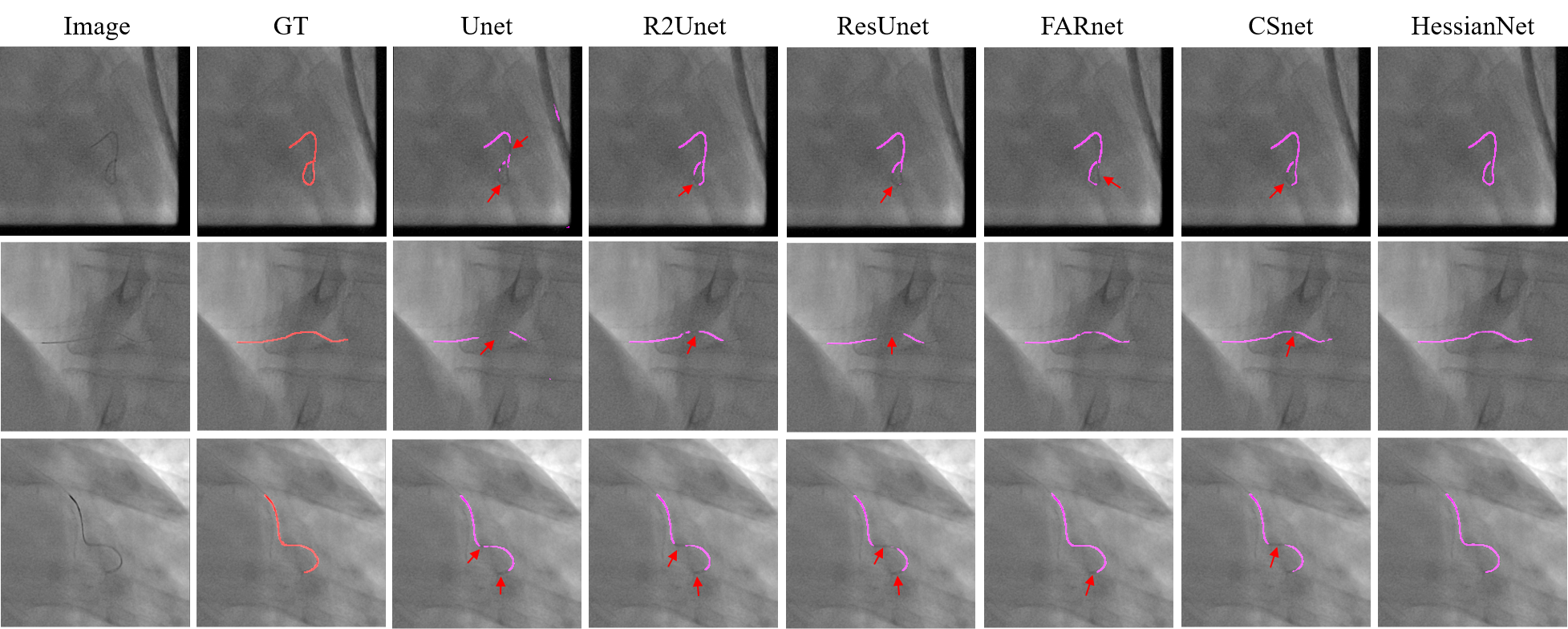}
   \end{center}
   \caption{Qualitative results of our and the state-of-the-art methods. Failure cases are highlighted with red arrows.}
   \vspace{2mm}
   \label{fig:qualitative_results} 
\end{figure}

 We compare our method to recent state-of-the-art segmentation models using the following metrics: Dice, Sensitivity, False Discovery Rate (FDR), and Hausdorff Distance (HD)~\cite{taha2015metrics}. Further, we evaluate the effect of including the proposed hessian layer and dual self-attention block, for which the results are summarized in Table~\ref{table:sota}. Meanwhile, qualitative examples of our and the baseline methods can be found in Fig~\ref{fig:qualitative_results}. From both the quantitative results and  the qualitative results, it is clear that our proposed method outperforms other methods. 
\begin{table*}[t!]
	\begin{center}
		\caption{Performance of our method, HessianNet, in comparison to the state of the art.}
		\vspace{-1mm}
		\smallskip
		\resizebox{\textwidth}{!}{	
		\setlength{\tabcolsep}{6pt}
			\begin{tabular}{lcccccccc}
			    \noalign{\smallskip}
			    \toprule
			     Method & U-Net \cite{ronneberger2015u} & R2UNet \cite{alom2018recurrent} & ResUnet \cite{zhang2018road} & FARNet \cite{zhou2020real} & CSNet \cite{mou2021cs2} & \vtop{\hbox{\strut HessianNet }\hbox{\strut w/o Attention}} & \vtop{\hbox{\strut HessianNet }\hbox{\strut w/o Enhance}} & HessianNet\\
				\noalign{\smallskip}
				\midrule
				 Dice & 0.8890 & 0.8939 & 0.8968 & 0.8981 & 0.8984 & 0.8987 & 0.8984 & \textbf{0.8990} \\
				 Sensitivity & 0.8840 & 0.8960 & 0.8908 & 0.8921 & 0.8909 & 0.8954 & 0.8970 & \textbf{0.8973} \\
				 FDR & 0.0987 & 0.1030 & 0.0923 & 0.1013 & \textbf{0.0897} & 0.0915 & 0.0930 & 0.0947 \\
				 HD & 0.4089 & 0.3488 & 0.3225 & 0.3362 & 0.3439 &0.3123 & 0.3126 & \textbf{0.3070} \\
				\noalign{\smallskip}
				\bottomrule
			\end{tabular}
		}
		\label{table:sota}
	\end{center}
	\vspace{1mm}
\end{table*}

 To further evaluate our methods' performance in comparison to the baselines, we provide an analysis with respect to the most difficult examples in our dataset. Based on the predicted segmentations of all methods, any image with a HD score larger than 0.1 will be regard as a difficult example. Table \ref{table:samples} summarizes our findings, which further indicates our proposed method is superior to these baseline methods and achieves the state of art.

\begin{table*}[t!]
	\begin{center}
		\caption{Performance of our method, on difficult examples, in comparison to the state of the art.}
		\vspace{-1mm}
		\smallskip
		\resizebox{\textwidth}{!}{	
		\setlength{\tabcolsep}{6pt}
			\begin{tabular}{lcccccccc}
			    \noalign{\smallskip}
			    \toprule
			     Method & U-Net \cite{ronneberger2015u} & R2UNet \cite{alom2018recurrent} & ResUnet \cite{zhang2018road} & FARNet \cite{zhou2020real} & CSNet \cite{mou2021cs2} & \vtop{\hbox{\strut HessianNet }\hbox{\strut w/o Attention}} & \vtop{\hbox{\strut HessianNet }\hbox{\strut w/o Enhance}} & HessianNet\\
				\noalign{\smallskip}
				\midrule
				 Dice & 0.7592 & 0.8137 & 0.8149 & 0.8198 & 0.8228 & 0.8270 & 0.8260 & \textbf{0.8292} \\
				 Sensitivity & 0.6932 & 0.7985 & 0.7767 & 0.8090 & 0.7992 & 0.8113 & 0.8115 & \textbf{0.8196} \\
				 FDR & \textbf{0.1382} & 0.1610 & 0.1356 & 0.1630 & 0.1417 & 0.1538 & 0.1554 & 0.1571 \\
				 HD & 2.2291 & 1.0787 & 0.9019 & 1.0454 & 0.7979 & 0.7116 & 0.6962 & \textbf{0.65520} \\
				\noalign{\smallskip}
				\bottomrule
			\end{tabular}
		}
		\label{table:samples}
	\end{center}
	\vspace{1mm}
\end{table*}

\begin{figure}[t!]
   \begin{center}
   \includegraphics[width=1.0\textwidth]{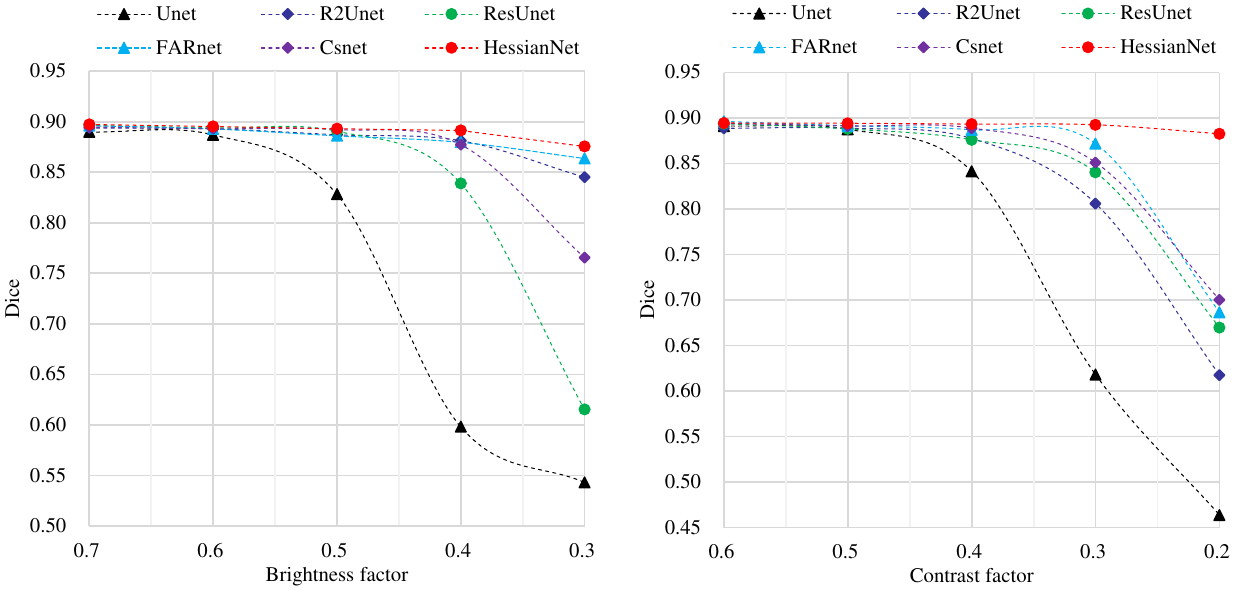}
   \end{center}
   \caption{Robustness of our and the baseline methods to variations in image brightness and contrast.}
   \label{fig:robustness} 
   \vspace{2mm}
\end{figure}

In addition, We analyse the robustness of our and the baseline methods to changes in image brightness and contrast during inference. Fig. \ref{fig:robustness} summarizes the dice score with respect to varying brightness and contrast ratios and shows that in comparison to our method the baselines are highly susceptible to such changes, resulting in a large drop in dice score.

\section{CONCLUSIONS}
\label{sec:conclusion}  % \label{} allows reference to this section
 We propose a two-stage framework for real-time guidewire tracking and segmentation. For the detection phase, an improved detector based on Yolov5s model is trained and a refinement module based on spatiotemporal constraint is proposed to robustly localize the guidewire and suppress false positive detections. For the segmentation phase, a novel and efficient network which integrates a hessian-based enhancement embedding module and a dual self-attention block is proposed to segment the guidewire from each cropped patch. Quantitative and qualitative evaluation results show that our method outperforms our baselines as well as the current state of the art and has a strong robustness for this task. Moreover, the whole system is designed to perform in real-time and achieves an inference rate of approximately 35 FPS on a NVIDIA GPU (Quadro RTX 6000). 

\section*{ACKNOWLEDGMENTS}
The project was funded by the Bavarian State Ministry of Science and Arts within the framework of the "Digitaler Herz-OP" project under the grant number 1530/891 02 and the China Scholarship Council (File No.202004910390).
% References
\bibliography{report} % bibliography data in report.bib
\bibliographystyle{spiebib} % makes bibtex use spiebib.bst
\end{document}